# DISCOVERY OF GOOD DOUBLE AND TRIPLE CIRCULANT CODES USING MULTIPLE IMPULSE METHOD

**ASKALI M.*, NOUH S., AZOUAOUI A. AND BELKASMI M.**

National School of Computer Science and Systems Analysis (ENSIAS), Mohammed V - Souisi University, Rabat, Morocco.
*Corresponding Author: Email- askali11@gmail.com

**Abstract-** The construction of optimal linear block error-correcting codes is not an easy problem, for this, many studies describe methods for generating good error correcting codes in terms of minimum distance. In a previous work, we have presented the multiple impulse method (MIM) to evaluate the minimum distance of linear codes. In this paper we will present an optimization of the MIM method by genetic algorithms, and we found many new optimal Double and Triple Circulant Codes (DCC & TCC) with the highest known parameters using the MIM method as an evaluator of the minimum distance. Two approaches are used in the exploration of the space of generators; the first is based on genetic algorithms, however the second is on the random search algorithm.

**Keywords-** Minimum distance, Error impulse Method, heuristic methods, Genetic algorithms, NP-hardness, Linear error correcting codes, Double and Triple Circulant Codes

## Introduction

The design of good codes is of fundamental importance in a communication system. Furthermore, finding good linear or nonlinear codes may affect the sphere packing problems in Euclidean spaces [1]. When the code rate is 1/2 or 1/3; Double or Triple Circulant Codes (DCC & TCC) have been an interesting family of codes with high minimum distances. It is still hard to determine the minimum distances of long binary codes as well as their asymptotic relative minimum distances [2,4]. Besides, Karlin [5] and Pless [3] found many good codes by systematic double circulant codes over GF(2) and GF(3) using quadratic residues respectively. In [6], Gaborit proposes a double circulant code scheme which generalizes the constructions of Karlin and Pless over any field and for any length n=pm, where p is an odd prime. Furthermore, Karlin considered binary circulant [3p+1, p+1] and [3p, p] codes using quadratic residues and nonresidues [5]. Recently artificial intelligence techniques were introduced to solve this problem. Among related works, one idea used Genetic Algorithms (GA) to design constant weight codes [7], another one used GA for searching the minimum distance of BCH code [8]. Lacan et al. [9] introduced Genetic algorithms in the search of optimal error correcting codes, and in [10], authors give new good DCC constructed by GA. Here, we propose two heuristic search methods such as Genetic Algorithm, and random search of DCC and TCC when we use the MIM method published in [11] in order to determine the minimum distance, and we give an improvement by introducing the multiple impulse error using genetic algorithm, which we call MIM-GA.

A table of the best known codes is regularly updated on site of code tables maintained by Markus Grassl [17]. For each pair of parameters n and k, this table contains the distance d of the best known code and its theoretical upper bound. In this paper, we are going to search for optimal error-Correcting double circulant codes, by the exploration of the space by Genetic Algorithms and random search. Besides, we will present a new optimization which reduces the complexity. Hence, we propose a technique based on heuristic methods to search of good DCC and TCC codes which have not been previously developed, at our knowledge, for this family of codes.

The paper is organized as follows: In the beginning, we give backgrounds about error correcting codes, and the manner of construction of DCC and TCC codes. Then we introduce genetic algorithms. After we give the implementing methods to evaluate minimum distance, and we improve the MIM method to search good DCC and TCC codes. In the last some experimental results and discussion are presented.

## Error Correcting Codes (ECC)

The In this section, we give the basics of error correcting codes, in particular we introduce the construction of Double and Triple Circulant error correcting codes (DCC & TCC).

### Error Correcting Codes

In all communication systems, the information transmitted is represented via a source code as the ASCII code, Huffman code, etc. This source-encoded information is then sent over a Channel, such as a telephone line, optical fiber, microwave link, etc. To have a reliable transmission the data are encoded again by using an error correcting code that enables the detection and the correction of possible errors introduced during the transmission of message [Fig-1]. Several modes of efficient coding are known: Hamming codes, Reed-Solomon codes, etc.

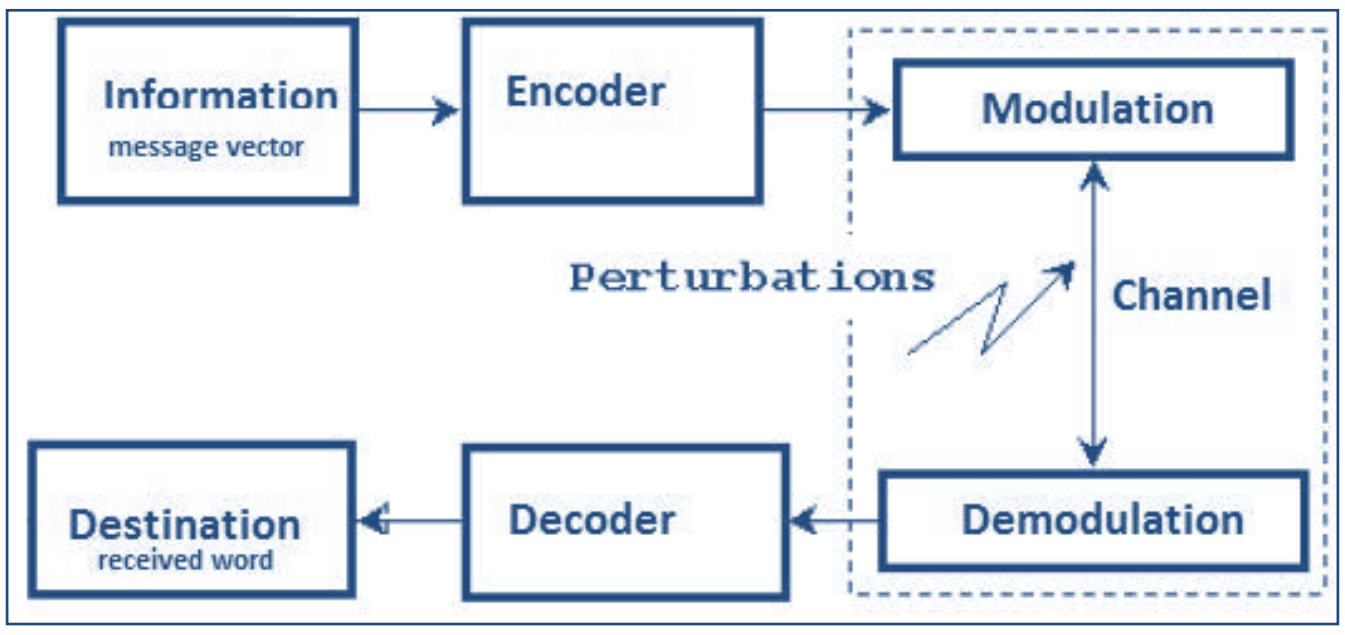

**Fig. 1-** A simplified model of communication system.

There exist principally two classes of error correcting Codes: convolutional codes and block codes. In this paper, we will focus on a special case of block codes: double and triple circulant codes (DCC & TCC). Let us explain more precisely the background of block code. Firstly, Let us suppose now that the information to transmit is a sequence of elements which take q possible values, where q is a power of the finite field of cardinal p, which is denoted by $F_q$. In general, transmitted symbols are binary; then q=2. Principle of block code is the following: the initial message is cut out into blocks of length k. The length of the redundancy is n-k and thus, the length of the transmitted blocks is n. Main blocks are linear codes. In this case, redundancy is computed in such a way as concatenation of information and of redundancy is an element of vector space (a code) of dimension k of $(F_q)^n$.

The operation of coding can be computed by multiplying the message (considered as a vector of length k) by a k*n systematic generator matrix of this vector space. Note that a generator matrix is called systematic if their k first columns form the identity matrix. For a giving code, there exists at most one systematic generator matrix. Some codes did not admit a systematic generator matrix. On the other hand, all these codes are equivalent (modulo a permutation of the positions) to a code which admits one. For a linear code, k and n are respectively called the dimension and the length of the code. The last parameter of a code is its minimum distance d. It is the smallest Hamming distance (the number of distinct components) between two codewords. As the considered codes are linear vector spaces, their minimum distance is also the Hamming weight (the number of nonzero components) of the codeword of smaller hamming weight. The correction capability (the maximum number of error that can be corrected per word of length n) of the code is then equal to t= (d-1)/2.

**Example 1:** Let C be a code of length 8 and dimension 4 over F2 characterized by its systematic generator G:

$$G=\begin{pmatrix}1\,0\,0\,0 & 1\,1\,1\,0\\ 0\,1\,0\,0 & 0\,1\,1\,1\\ 0\,0\,1\,0 & 1\,0\,1\,1\\ 0\,0\,0\,1 & 1\,1\,0\,1\end{pmatrix}$$

The transmitted codeword corresponding to the message (1101) is: (1101) x G = (1101 0100).

The set all possible messages (information vectors), the corresponding set of codewords and their weight are shown in [Table-1].

*Table 1- Codewords of a code C (8, 4, 4)*

| Message | Codeword | Weight |
|---|---|---|
| 0 | 0000 0000 | 0 |
| 1 | 0001 1101 | 4 |
| 10 | 0010 1011 | 4 |
| 11 | 0011 0110 | 4 |
| 100 | 0100 0111 | 4 |
| 101 | 0101 1010 | 4 |
| 110 | 0110 1100 | 4 |
| 111 | 0111 0001 | 4 |
| 1000 | 1000 1110 | 4 |
| 1001 | 1001 0011 | 4 |
| 1010 | 1010 0101 | 4 |
| 1011 | 10111000 | 4 |
| 1100 | 1100 1001 | 4 |
| 1101 | 1101 0100 | 4 |
| 1110 | 1110 0010 | 4 |
| 1111 | 1111 1111 | 8 |

## Double Circulant Codes (DCC)

An r x r matrix A= $[Aij]_{0\le i,j\le r-1}$ over an alphabet F is called circulant if $A_{ij}=A_{0,j-i}$ for all 0≤ i, j≤ r-1, where indices are computed modulo r. Let the notation [n, k, d] stand for a k-dimensional linear code of length n and minimum distance d over F, and let r=n-k denotes the redundancy of the code. An [n=2r, k=r, d] linear code over F is called a double circulant if its generated matrix G=[I A], where A is an r x r circulant matrix over F[2,p. 497]. Double circulant codes have been discussed extensively in the literature [2] and a subclass of DCC approaches the Gilbert-Varshamov bound [14]. We are interested by this family of codes because it contains good codes with a maximum minimum distance.

Consider a double circuant code C generated by a matrix [I | A] where A= $[a_{ij}]_{0\le i,j\le r-1}$. The vector ($a_0$,…, $a_{r-1}$) is called the header of generator matrix of DCC and I is the identity matrix. With every ($a_0$, …, $a_{r-1}$) є $F_r$, we associate a matrix A= $[A_{ij}]_{0\le i,j\le r-1}$ as the [Eq-1]:

$$A=\begin{pmatrix} a_0 & a_1 & a_2 & \cdots & a_{r-4} & a_{r-3} & a_{r-2} & a_{r-1}\\ a_{r-1} & a_0 & a_1 & \cdots & a_{r-5} & a_{r-4} & a_{r-3} & a_{r-2}\\ \vdots & \vdots & \vdots & \cdots & \vdots & \vdots & \vdots & \vdots\\ a_1 & a_2 & a_3 & \cdots & a_{r-3} & a_{r-2} & a_{r-1} & a_0 \end{pmatrix} \quad (1)$$

Where ($a_0$, $a_1$,…………….,$a_{r-1}$) is the header of the A circulant matrix. Each header corresponds to exactly one DCC.

**Example 2:** Let C (18, 9, 6) a double circulant where her header generator matrix is (011101001) we have the generator matrix G is as the following:

$$G=\begin{pmatrix} 100000000 & 011101001\\ 010000000 & 101110100\\ 001000000 & 010111010\\ 000100000 & 001011101\\ 000010000 & 100101110\\ 000001000 & 010010111\\ 000000100 & 101001011\\ 000000010 & 110100101\\ 000000001 & 111010010 \end{pmatrix}$$

There exists other subfamily of Double Circulant codes like the bordered double Circulant Codes witch is represented by the form described in [Fig-2].

$$G = \left[\begin{array}{c|c|c|c} 1 & & 0 & \\ \vdots & I_p & \vdots & B \\ 1 & & 0 & \\ \hline 0 & 0 \; \dots \; 0 & 1 & 1 \; \dots \; 1 \end{array}\right]$$

**Fig. 2-** Bordered form of Double Circulant Code

*B = Circulant matrix presented by one header*

*$I_p$ = Identity matrix p*p.*

### Triple Circulant Codes (TCC)

The Let the notation [n, k, d] stand for a k-dimensional linear code of length n and minimum distance d over F, and let r=n-k denotes the redundancy of the code. An [n=3r, k=r, d] linear code over F is called a triple circulant if its generated matrix G= [I A B], where A and B are two r x r circulant matrices. We are interested by this family of codes because by experimentation we found good codes with a maximum minimum distance.

Consider a Triple circuant code C generated by a matrix [I | A | B] where A= $[a_{ij}]_{0\le i,j\le r-1}$ and B= $[b_{ij}]_{0\le i,j\le r-1}$.

The vectors ($a_0, a_1,\dots, a_{r-1}$) and ($b_0, b_1,\dots, b_{r-1}$) are called the headers of generator matrix of TCC and I is the identity matrix. With every header ($a_0, a_1,\dots, a_{r-1}$) є $F_r$, we associate a matrix A= $[A_{ij}]_{0\le i,j\le r-1}$ and every header ($b_0, b_1,\dots, b_{r-1}$) є $F_r$, we associate a matrix B= $[b_{ij}]_{0\le i,j\le r-1}$ as the [Eq-2] and [Eq-3]:

$$A = \begin{pmatrix} a_0 & a_1 & a_2 & \cdots a_{r-4} & a_{r-3} & a_{r-2} & a_{r-1} \\ a_{r-1} & a_0 & a_1 & \cdots a_{r-5} & a_{r-4} & a_{r-3} & a_{r-2} \\ \vdots & \vdots & \vdots & \cdots \vdots & \vdots & \vdots & \vdots \\ a_1 & a_2 & a_3 & \cdots a_{r-3} & a_{r-2} & a_{r-1} & a_0 \end{pmatrix} \qquad (2)$$

$$B = \begin{pmatrix} b_0 & b_1 & b_2 & \cdot\cdot b_{r-4} & b_{r-3} & b_{r-2} & b_{r-1} \\ b_{r-1} & b_0 & b_1 & \cdot\cdot b_{r-5} & b_{r-4} & b_{r-3} & b_{r-2} \\ \vdots & \vdots & \vdots & \cdots \vdots & \vdots & \vdots & \vdots \\ b_1 & b_2 & b_3 & \cdot\cdot b_{r-3} & b_{r-2} & b_{r-1} & b_0 \end{pmatrix} \qquad (3)$$

Where

($a_0, a_1,\dots\dots\dots\dots,a_{r-1}$) is the header of the circulant matrix A, and ($b_0, b_1,\dots\dots\dots\dots,b_{r-1}$) is the header of the circulant matrix B.

**Example 3:** Let C (27, 9, 6) a Triple circulant code, where its headers of the generator matrix G are a=(011101001) and b=(110110110), G is as the following :

$$G = \begin{pmatrix} 100000000 & 011101001 & 110110110 \\ 010000000 & 101110100 & 011011011 \\ 001000000 & 010111010 & 101101101 \\ 000100000 & 001011101 & 110110110 \\ 000010000 & 100101110 & 011011011 \\ 000001000 & 010010111 & 101101101 \\ 000000100 & 101001011 & 110110110 \\ 000000010 & 110100101 & 011011011 \\ 000000001 & 111010010 & 101101101 \end{pmatrix}$$

### Genetic Algorithms

Before, Genetic Algorithms (GA) was first proposed by John Holland's, as a means to find good solutions to problems that were otherwise computationally intractable. Holland's schema theorem [12], and the related building block hypothesis, provided a theoretical and conceptual basis for the design of efficient GA. It also proved straight forward to implement GA due to their highly modular nature. As a consequence, the field grew quickly and the technique was successfully applied to a wide range of practical problems in science, engineering and industry. GA theory is an active and growing area, with a range of approaches being used to describe and explain phenomena not anticipated by earlier theory. In tandem with this, more sophisticated approaches for directing the evolution of a GA population are aimed at improving performance on classes of problem known to be difficult for GA, [12]. The development and success of GA contributed greatly to a wider interest in computational approaches based on natural phenomena. It is now a major stand of the wider field of computational intelligence, which encompasses techniques such as neural networks, and artificial immunology. Genetic algorithms are search methods that can be used for both solving problems and modelling evolutionary systems. Since it is heuristic (it estimates a solution), GA differs from other heuristic methods in several ways. The most important difference is that it works on a population of possible solutions; while other heuristic methods use another important difference is that GA is not a deterministic but a probabilistic one.

A genetic algorithm is defined by [Fig-3]:

Individual or chromosome: a potential solution of the problem, it's a sequence of genes.

- Population: a set of points of the research space.
- Environment: the space of research.
- Fitness function: the function to maximize / minimize.

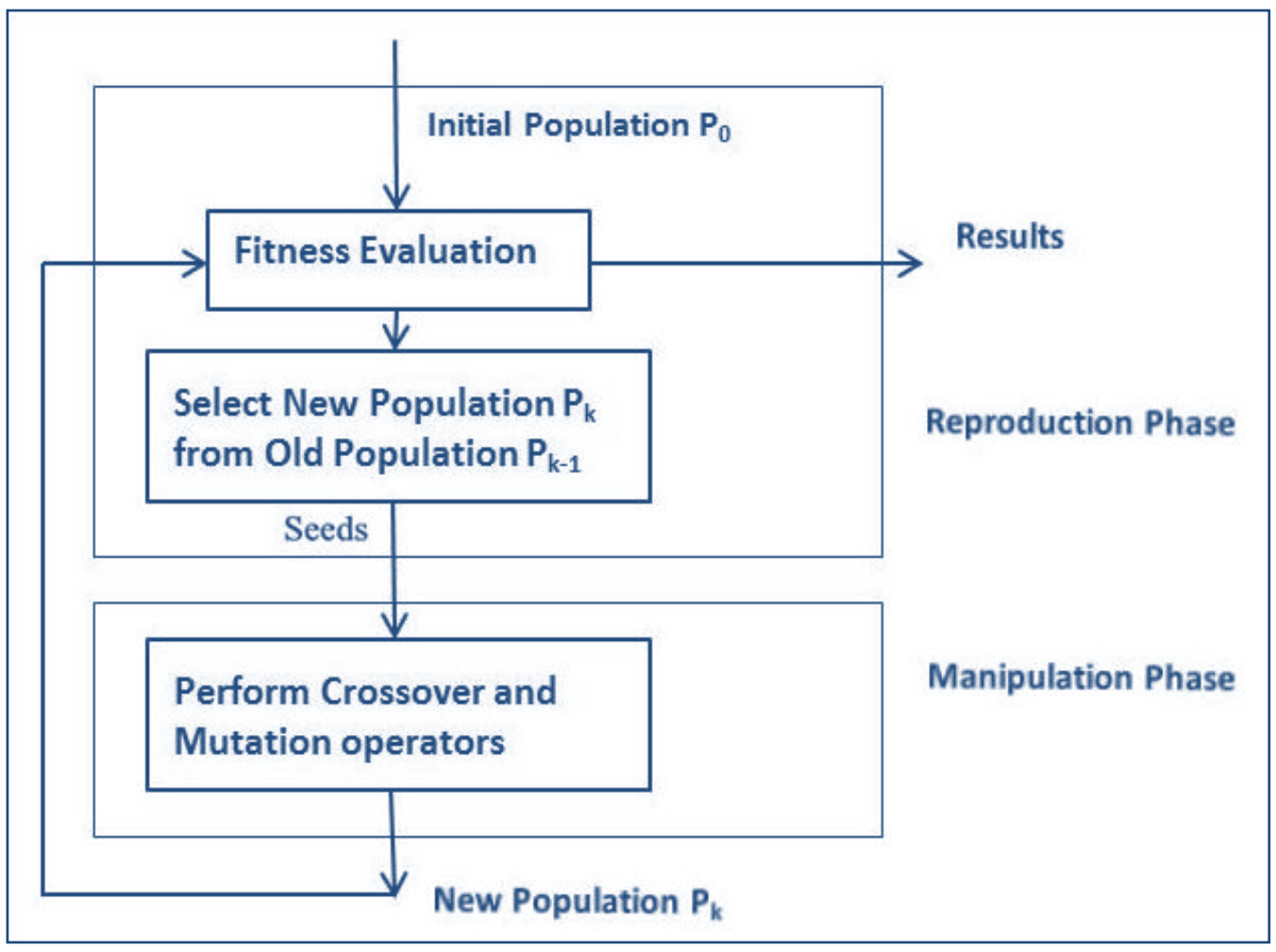

**Fig. 3-** The Basic Structure of Genetic Algorithm

Encoding of Chromosomes: it depends on the treated problem, the famous known schemes of coding are:

Binary encoding, permutation encoding, value encoding and tree encoding.

The Stochastic Operators are :

- Selection: replicates the most successful solutions found in a population at a rate proportional to their relative quality.

- Crossover: Decomposes two distinct solutions and then randomly mixes their parts to form novel solutions.
- Mutation: Randomly perturbs a candidate solution. In the selection process, some individuals are selected to be copied into a tentative next population. Individual with higher fitness value is more likely to be selected. The selected individuals are altered by the mutation and crossover and form a new population of solutions. The GA is simple yet provides an adaptive and robust optimization methodology [13].

### Evaluators of the Minimum Distance

In this section, we give some methods previously used to estimate the minimum distance of error correcting codes; Multiple Impulse Method (MIM) and Genetic algorithm evaluators used by authors for some linear block codes [11], and Chen's Method for Cyclic codes [15].

### Multiple Impulse Method (MIM)

This method produces a tight minimum distance based on true (low-weight) codewords found by a fine-tuned local search. the principal is injecting a noise iteratively in a multiple random positions, on the contrary of error impulse method used by Berrou in [18] where the minimum distance is the magnitude of the noise that we inject to the 'all zero" codeword' in one position. In the MIM method the decoded word in the output of the Soft-In OSD decoder will be mostly near than the "all-zero" codeword, and the minimum distance of the code will be the minimum weight of the decoded words. The MIM algorithm described in [11] is as follow:

### MIM Algorithm

*Inputs:* G, the generator matrix

We assume that $d_t$ is in the range $[d_0,d_1]$ where $d_0$ and $d_1$ are two integers. Then $d_t$ can be determined as follows:

*Step 1*: set Amin=d1+0.5 and dt =n - k 1;nb_test.

*Step 2*: For i=1 to nb_test

*Step 2.1*: A = d0 – 0.5;

*Step 2.2*: Set [( $\tilde{x} = x$)=TRUE];

*Step 2.3*: While[( $\tilde{x} = x$) = TRUE] && [A ≤ Amin-1.0 ]

*Step 2.3.1*: A = A + 1.0

*Step 2.3.2*: For nb_error=error_max to 1

*Step 2.3.3*: Subdivide A randomly on nb_error positions

*Step 2.3.3.1*: Y ←x (after modulation) +A

*Step 2.3.3.2*: OSD decoding of Y $\rightarrow \tilde{x}$

*Step 2.3.3.3*: If (weight( $\tilde{x}$) ≤ $d_t$ ) then $d_t$ = weight( $\tilde{x}$ )

*Step 2.3.3.4*: If( $\tilde{x} \neq x$) then [( $\tilde{x} = x$)=TRUE]

End for

End while

Step 2.4: Amin=A;

End for

Output: $d_t$ is the minimum distance

### Chen's Method

A general result for cyclic codes, due to Chen [15] is the following:

**Theorem 1:** Let c be a codeword of a cyclic [n, k] code. Suppose that the Hamming weight of c is equal to **w**. Then there exists a cyclic shift of c with exactly $r = \left\lfloor \frac{kw}{n} \right\rfloor$. Nonzero coordinates among its first k coordinates.

Some improvements of this theorem were obtained by Voloch [16]. With Chen's theorem, it is then possible to look for code words of weight w in a cyclic [n, k] code, given the systematic encoding matrix of the code.

Let $G = [I_k\ M]$ be such a matrix where $I_k$ is the identity matrix of size k. Then, if a codeword of Hamming weight w exists, it can be obtained by the linear combination of r rows of G. Chen's theorem gives then an explicit way to enumerate low weight codewords, provided that the number of combinations is not too large.

More precisely, if we denote $a_{ij}$ for $0 \le i, j \le k$ the binary element at row i and column j of M, then for a given r value, Chen's algorithm amounts to enumerating vectors of the form:

$$\bigoplus_{1 \le i_1 \le i_2 \le \ldots\ldots \le i_r \le k} \left[ m_{i_1,1}, m_{i_2,2}, \ldots\ldots m_{i_r,r} \right]$$

Where $\oplus$ denotes column wise modulo 2 operation. For each of these vectors, the Hamming weight is computed. If $\mu$ is the minimum value of the obtained Hamming weights, then the minimum Hamming weight of codewords obtained by linear combination of r rows of G is then $r + \mu$. Doing this, for each possible w (and r) value, it is then possible to determine the minimum distance of the cyclic code generated by matrix G. In this work we use Chen's Theorem to minimize the computing of the minimum distance used in the genetic algorithm presented in [10]. And we will focus on a special case of block codes: double and triple circulant codes.

### Genetic Method to Find the Minimum Sistance

The steps of the algorithm are organized as follow:

*Step 1*: randomly generate an initial population

Seed uniformly, randomly the initial population with a $N_i$, and where each individual is a word of length k with a random weight. We initiate the number of generation $N_g$ to 1.

*Step 2*: while ($N_g < N_{gmax}$) do

*Step 2.1*: Compute the fitness of each individual in the population

An individual i represents an information vector of k bits which is encoded by the generator code to an n-bit code vector. The fitness is the weight of the encoded individual if this last is different to zero otherwise, the fitness is equal to n as a maximum value.

$$f \leftarrow weight\ (coding\ individual)$$

*Step 2.2*: Sort population in increasing order of fitness

*Step 2.3*: select the best $N_e$ individuals in the intermediate population

*Step 2.4*: For i=$N_e$ to $N_i$

*Step 2.4.1*: tournament select of two parents $p_1$ and $p_2$ for reproduction

*Step 2.4.2*: If ( rand_value < $p_c$ ) { Cross $p_1$ and $p_2$ to generate $ch_1$ and $ch_2$; Mutate $ch_1$ and $ch_2$ and introduce them in the next population} Else introduce $p_1$ or $p_2$ into the next population with equal probability.

End For

*Step 2.5*: Let currbest=fittest of the intermediate population. If(fitness (best) < fitness(currbest)) best=currbest

End while

*Step 3*: output best

### Description of the Algorithm

In Step 2.4.1, we use the tournament selection, in that only one of two possible parents is preserved to reproduce two children whose will be inserted in the next generation.

Step 2.4.2, in this variant, the crossover operation depends on pc, and it is done before the mutation step which is done bit-wise on offspring with probability pm. In case of no-cross we insert the two initials parents in the next generation. We have used three strategies of crossover: a single crossover point, two point crossover, and uniform crossover. The two-point Crossover that randomly selects two crossover points within a chromosome then interchanges the two parent chromosomes between these points to produce two new offspring [Fig-4]. The Uniform Crossover uses a fixed mixing ratio between two parents. Unlike one- and two-point crossover, the Uniform Crossover enables the parent chromosomes to contribute the gene level rather than the segment level. An example of this operation is depicted in [Fig-5].

### An Optimization of the Multiple Impulse Method by Genetic Algorithms (MIM-GA)

From the MIM method presented above, we can pose the key question bellow: Which values of the parameters *A* and *nb_error* are good in terms of the minimization which they make on the weight of the decoded word by the OSD decoder ? In other words, which is the good impulse?

In this work, we will use a genetic algorithm to find these appropriate values. Consequently, the MIM-GA method will be works as follows:

### MIM-GA Algorithm

Inputs:
- The population size $N_{ind}$
- The maximum number of generations $N_{gm}$
- The crossover probability $p_{cr}$
- The mutation probability $p_{mu}$
- The mutation amplitude r
- The assumed interval $[d_0,d_1]$
- The number of positions nb_error

Outputs: $d_t$

Begin

*Step 1*: Seed uniformly, randomly the initial population of $N_{ind}$ individuals, and where each individual is a word of n genes. Each gene is a reel value. These words are obtained by subdividing some random values (A between $d_0$ and $d_1$) on nb_error positions. At each individual, the modulated zero word is added. We initiate the number of generation $N_g$ to 1 and $d_t$ to n.

*Step 2*: While($N_g < N_{gm}$) do

*Step 2.1*: Compute the fitness of each individual in the population

*Step 2.1.1*: OSD decoding of the individual $\rightarrow \tilde{x}$

*Step 2.1.2*: f←weight( $\tilde{x}$ )

*Step 2.1.2*: if(f=0) then f←n

*Step 2.1.3*: If($f \le d_t$) then dt ← f

*Step 2.2*: Sort population in increasing order of fitness

*Step 2.3*: insert the best individuals in the current population

*Step 2.4*: For i=2 to Nind

*Step 2.4.1*: Randomly select two parents $p_1$ and $p_2$ for reproduction

*Step 2.4.2*: If ( rand_value < $p_{cr}$ )

{ Cross $p_1$ and $p_2$ to generate ch;
Mutate ch according to pmu and r;
introduce ch in the current population }

Else insert $p_1$ or $p_2$ into the next population with equal probability.

End For

End while

*Step 3*: output $d_t$

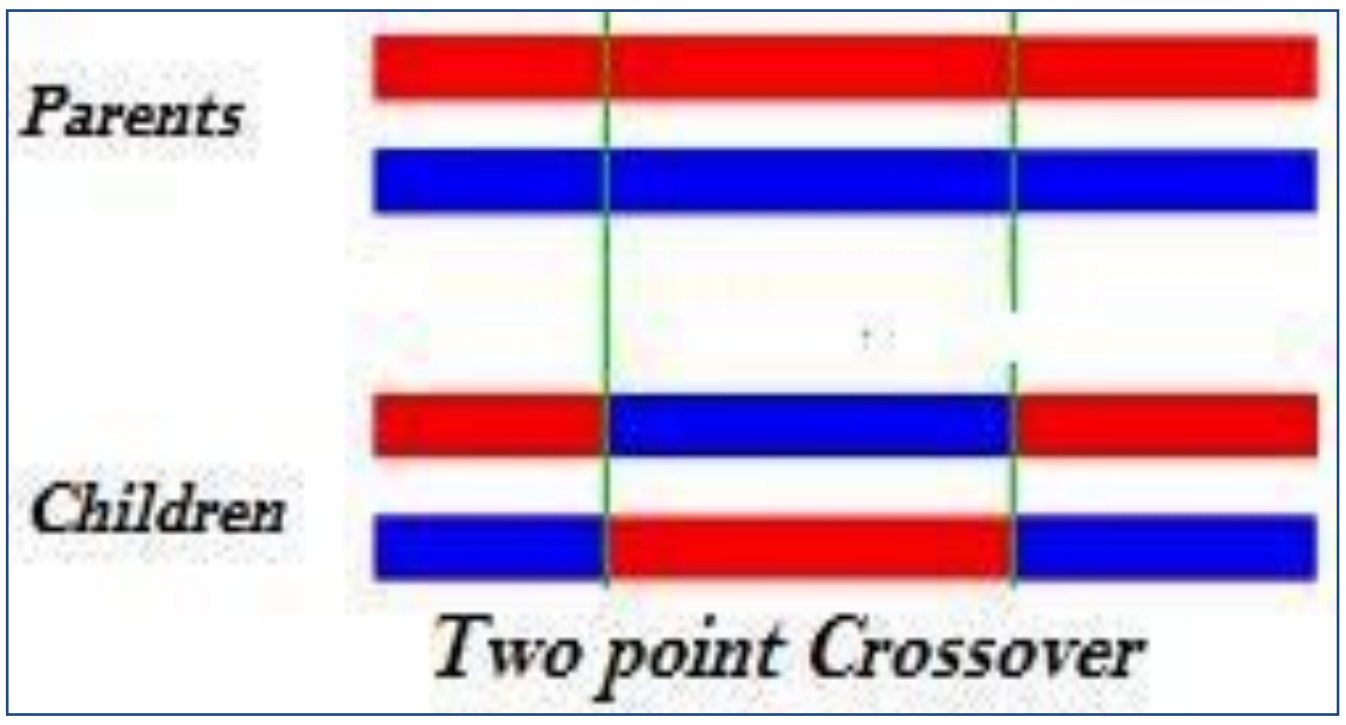

**Fig. 4-** Two-point Crossover structure.

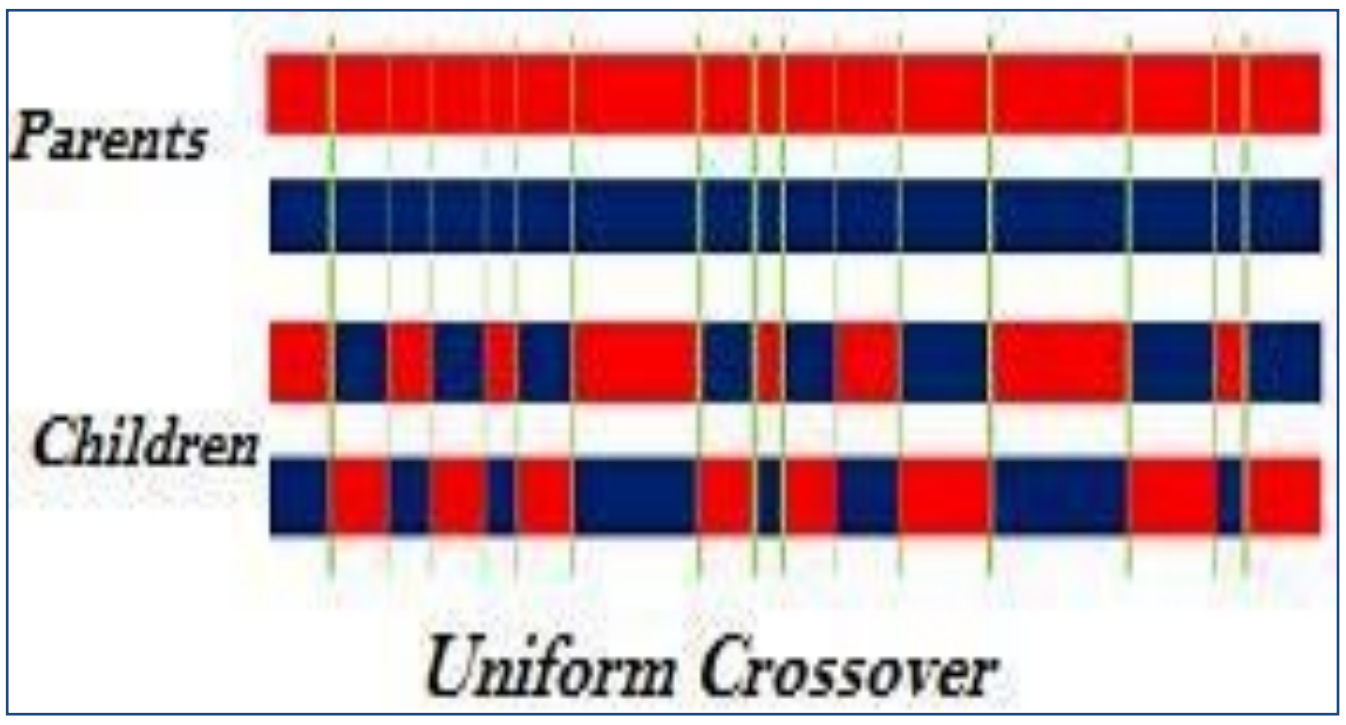

**Fig. 5-** Uniform Crossover structure.

### Comparison between MIM and MIM-GA Algorithms

In order to compare the run times of MIM-GA and MIM algorithms, we fixe all the parameters of MIM-GA at the values outlined in the [Table-2] below:

*Table 2- Parameters of implementation of Genetic Algorithms*

| Parameter of GA | value |
|---|---|
| $N_{ind}$ | 10 |
| $N_{gm}$ | 10000 |
| $p_{cr}$ | 0.95 |
| $p_{mu}$ | 0.05 |
| r | 0.1 |
| Crossover type | One point |
| Order of the OSD decoder | 2 |
| Mutation of a gene g | g←g±r |

The [Table-3] and [Table-4] below give the run time in seconds of the MIM and MIM-GA algorithms executed for some QR and BCH codes in the same computer, This run time is equal to the beginning of the algorithm and the moment of finding the lowest weight codeword.

*Table 3- Comparaison of the Run time of MIM and MIM-GA for some QR codes*

| QR Code | Distance by MIM | Distance by MIM-GA | Run time of MIM(s) | Run time of MIM-GA(s) |
|---|---|---|---|---|
| n=233 k=117 | 25 | 25 | 198 | 11 |
| n=241 k=121 | 31 | 31 | 835 | 218 |
| n=257 k=129 | 33 | 33 | 5851 | 980 |
| n=281 k=141 | 35 | 35 | 40435 | 94 |
| n=313 k=157 | 45 | 45 | 51498 | 7860 |
| n=337 k=169 | 51 | 51 | 45539 | 8239 |

*Table 4- Comparaison of the Run time of MIM and MIM-GA for some BCH codes*

| BCH(n,k,d-design) | Distance by MIM | Distance by MIM-GA | Run time of MIM(s) | Run time of MIM-GA(s) |
|---|---|---|---|---|
| BCH(127,64,21) | 21 | 21 | 10 | 1 |
| BCH(127,57,23) | 23 | 23 | 1 | 1 |
| BCH(127,50,27) | 27 | 27 | 4 | 1 |
| BCH(255,71,59) | 61 | 61 | 530 | 101 |
| BCH(255,79,55) | 55 | 55 | 631 | 5 |
| BCH(255,87,53) | 53 | 53 | 5915 | 157 |
| BCH(255,91,51) | 51 | 51 | 7617 | 13 |
| BCH(255,115,43) | 43 | 43 | 8283 | 27 |
| BCH(255,123,39) | 39 | 39 | 7098 | 14 |
| BCH(255,131,37) | 37 | 37 | 7570 | 9 |
| BCH(255,139,31) | 31 | 31 | 7051 | 232 |
| BCH(255,147,29) | 29 | 29 | 4626 | 19 |
| BCH(255,155,27) | 27 | 27 | 4177 | 219 |
| BCH(255,163,25) | 25 | 25 | 2612 | 1391 |
| BCH(255,171,23) | 23 | 23 | 2847 | 247 |
| BCH(255,179,21) | 21 | 21 | 1653 | 9 |
| BCH(255,187,19) | 19 | 19 | 65 | 5 |
| BCH(255,191,17) | 17 | 17 | 1198 | 3 |
| BCH(255,199,15) | 15 | 15 | 384 | 1 |

The Table-3] and [Table-4] clearly show that the MIM and MIM-GA algorithms give the same value of the estimated minimum distance. However, the run time of MIM-GA is much reduced comparing to the MIM version.

### Proposed Methods to find Good DCC and TCC

In order to find good DCC and TCC, we propose two approaches, in the first we use genetic algorithms (GA variant) and in the second we search randomly good headers (RS variant).

### Genetic Variant using MIM

All simulations of the GA variant used to discover good codes were made with default GA parameters outlined in the [Table-5].

The Algorithm of the GA Variant is described as follow

1. Generate an initial population, of $N_i$ individuals, each individual is a word of weight $w^3$ upper bound-1 where its length is equal to k for DCC, and it is equal to 2k for TCC.

*Ng ← 0; fix* the value of Ngmax

2. Make evolve the population:

- while (Ng<Ngmax):
- Compute the fitness by MIM of individuals:
- fitness (individual) =d_MIM(individual)
- Sort the population by decreasing order of the fitness.
- m ← fitness (best individual)
- Copy the best $N_e$ individuals (of big fitness) in the intermediate population.
- For i=$N_e$ to $N_i$ :

  ∗ Select a couple of parents (p1,p2) of individuals among the better.

  ∗ Generate a random number x: 0 ≤x≤1

  ∗ if x < pc then:

  a) Cross p1 and p2 for generate ch1 and ch2

  b) Mute ch1 and ch2

  ∗ Among ch1, ch2 select the word c' of the biggest fitness and insert it in the intermediate population.
- *Ng ← Ng + 1.*

*Table 5- Parameters of implementation of GA variant.*

| Parameter of GA | value |
|---|---|
| Probability of Crossover | 80% |
| Probability of Mutation | 2% |
| Crossover Type | 2-point |
| Selection type | Tournament |
| Tournament size | 2 |
| Generation Number | 75 |
| Individuals Number | 10000/1000 |

### Good DCC found by Genetic Variant

The [Table-6] summarizes the good DCC codes that we found by the genetic variant, using the multiple impulse method. Where, we denote by LB the lower bound, and by UP, the Upper bound of the minimum distance of a given parameters of a linear code.

*Table 6- Good DCC codes obtained by GA variant*

| DCC | Binary Header of a good DCC | LB | $d_{MIM}$ | UB |
|---|---|---|---|---|
| C(202,101) | 101010101100010001100101100101001101000<br>0001101100000000010000010011111000011001<br>01101011100101001001001 | 28 | 30 | 46 |
| C(256,128) | 0000111111000101111011100011011111111110<br>100110101100010101001001001011100101110<br>000010001110011101001110111101100010010<br>11111111001 | 38 | 38 | 58 |
| C(190,95) | 011100011010101111101110010010010101000<br>101100001100100011000010001100101110110<br>00000100000111011 | 27 | 28 | 44 |

### Random Search Variant using MIM

The algorithm of the random search (RS) variant using multiple impulse method, we try to discover good codes with best parameters. The algorithm is described as follow:

Inputs: k, n, Max

For i= 1 to Max

  generate randomly the header of length k

  Generate the Generator matrix G related to $h_1$.

  Evaluate the minimum distance d of the Code generated by G using MIM

  If(d ≥ Lower Bound) save the code

End For

Outputs: list of good codes

### Good DCC found by Random Search Variant

The [Table-7] summarizes the good DCC codes that we found by random search variant, using the multiple impulse method and validated by the chen's method.

### Good TCC found by Random Search Variant

The [Table-8] summarizes the good TCC codes that we found by random search variant, using the multiple impulse method and validated by the chen's method.

*Table 7- Good DCC found using random search variant and validated by the Chen's method*

| DCC | Binary Header of a good DCC | LB | $d_{MIM}$ | UB | by Chen |
|---|---|---|---|---|---|
| C(140,70) | 110100001000101011010001101111101110 1110110111111011110011100010100110 | 22 | 22 | 32 | 22 |
| C(146,73) | 0010010111010100011000000011101010011 1110011001101001010001100101000010110 | 22 | 22 | 33 | 22 |
| C(146,73) | 1100011011110010111100100000011101000 111001001101011100111100110111110011 | 22 | 22 | 33 | 22 |
| C(160,80) | 1110011010110011011011100100011011100 0110111100101111001000000111010001110 010011 | 24 | 24 | 36 | 24 |
| C(190,95) | 0111000001101111100011001110110000100 1001001010010001110011011111111001100 1100000011000011000001 | 27 | 28 | 44 | 28 |
| C(156,78) | 0111010100001100000110100111001100010 1011010101000000010011101101010110100 1110 | 22 | 22 | 36 | 22 |
| C(156,78) | 0100000100010001100100110000111011001 0001110100001001001001101100011101011 1110 | 22 | 22 | 36 | 22 |
| C(156,78) | 1101110011111011100010111001010101001 1101110111111111111010010011000011111 0010 | 22 | 22 | 36 | 22 |
| C(192,96) | 1011000010011000100100010101100100111 0111010001011110000000111110000101010 1 0011100110101011100010 | 28 | 28 | 44 | 30 |
| C(192,96) | 0001000110001010010111001111001110000 0000010111110011101101010110001001010 1110011100100011100111 | 28 | 28 | 44 | 30 |
| C(192,96) | 1100001001010001010111011011111010001 0101010001110111001100011010001111000 1010100001100100101011 | 28 | 28 | 44 | 30 |
| C(188,94) | 1011110101100110011100010001100001101 0001111000111000111101010000100001001 0101101011000011010 0 | 26 | 27 | 43 | 27 |
| C(188,94) | 0100001110110000010111011010001010111 0011111110001010110000011001100101101 0111001101011101001 1 | 26 | 28 | 43 | 30 |
| C(188,94) | 0000100101001010011110000100111101100 1001000111011010011001001111000110100 100101010111110011 00 | 26 | 26 | 43 | 30 |
| C(192,96) | 0000001000001000101101001010010110011 1111011111011000001001110110001101011 0101110001010011110000 | 28 | 28 | 44 | 30 |
| C(188,94) | 0111100101110100110010001010001110011 1011101010001100010111100000110101001 00000001111110000111 | 26 | 27 | 43 | 29 |
| C(188,94) | 1111000100111011111110101011000110011 1100011000110101101000010011101111001 1100101001100010101 1 | 26 | 27 | 43 | 28 |
| C(188,94) | 1010111110001101001110111111011001100 1000011010010101011010111111110000000 0000010110000101010 1 | 26 | 27 | 43 | 30 |

*Table 8- Good TCC found using random search Algorithm and validated by the Chen's method*

| TCC | Binary Headers of a good TCC | LB | $d_{MIM}$ | UB | by Chen |
|---|---|---|---|---|---|
| C(36,12) | a= 001010011011 b= 100100000111 | 12 | 12 | 12 | 12 |
| C(42,14) | a= 10010011110000 b= 00100001010111 | 13 | 13 | 14 | 13 |
| C(45,15) | a= 110010111000110 b= 101000100110110 | 14 | 14 | 15 | 14 |
| C(54,18) | a= 100000010001010010 b= 000001011101110111 | 16 | 16 | 18 | 16 |
| C(57,19) | a= 1110101000110110 10 b= 0001001011000011 00 | 16 | 16 | 19 | 16 |
| C(102,34) | a=0101011111000000000101011001 1000100 b=1000011001010011110000001 10 0010010 | 24 | 24 | 32 | 24 |
| C(144,48) | a=1000100110001001000100111 11 00000000001010101 1001 b=011100100000011000010010010 010001100000110000010 | 32 | 32 | 45 | 32 |
| C(186,62) | a=0111110111000001011000111 10 0100110111001100011110111 1000 101101 b=1111110100100110011111011 11 1110001010110000100101100010 1 110110 | 38 | 40 | 58 | 40 |
| C(192,64) | a=1011110110011001111110010 10 0000011001011000110010101 1101 11101110 b=0110000101001011100010110 11 1001001111101100011101101 1111 10011011 | 41 | 42 | 60 | 44 |
| C(192,64) | a=1010011100110100011011101 10 0110011100011111111010000 1011 01110111 b=1100001010100111110011110 01 1011011101100110000101001 0101 10110000 | 41 | 42 | 60 | 42 |
| C(204,68) | a=1010011111100101111110111 10 1110000000000110001001011 10111 001101111010 b=1000011100010010001011010 01 1101000011010100001100110 0001 011000100000 | 41 | 42 | 62 | 48 |
| C(204,68) | a=0011100110011001101101101 01 1100100111110001000100010 0111 000000110010 b=1000011000111111001101101 10 0000011001101101011010110 1011 010101001000 | 41 | 43 | 62 | 47 |
| C(204,68) | a=1110111010000001001100001 100 0001110000100011111111110 010 110101101010 b=0010111010101111111111101 01 0110101111100010100110101 0010 001101100100 | 41 | 42 | 62 | 46 |
| C(204,68) | a=1000010000101010000100101 00 0000111110110110111101101 0010 000101010001 b=0101100111110101000011110 01 0110011001001110000001000 1000 100001000010 | 41 | 44 | 62 | 50 |

## Conclusion

In this paper, we have improved the MIM method in terms of its run time and we have, proposed two approaches to search good DCC and TCC. These techniques have given a high performance based on the presented integration of MIM method. Our results show that the MIM method explained in this work lead to good DCC and TCC, and we have found some codes in this family with a higher minimum distance than the lower bound of a given length and dimension. Most importantly, the technique proposed is useful also to be applied to deal with other type of codes especially non-binary ones.